%% file: main.tex
\documentclass[%
 reprint,
 amsmath,amssymb,
 aps,
]{revtex4-2}

\usepackage{dcolumn}% Align table columns on decimal point
\usepackage{bm}% bold math
\usepackage[dvipsnames]{xcolor}       % colors
\usepackage[utf8]{inputenc} % allow utf-8 input
\usepackage[T1]{fontenc}    % use 8-bit T1 fonts
\usepackage{hyperref}       % hyperlinks
\usepackage{url}            % simple URL typesetting
\usepackage{booktabs}       % professional-quality tables
\usepackage{diagbox}        % diagonal tables
\usepackage{amsfonts}       % blackboard math symbols
\usepackage{nicefrac}       % compact symbols for 1/2, etc.
\usepackage{microtype}      % microtypography
\usepackage{lipsum}
\usepackage{fancyhdr}       % header
\usepackage{graphicx}       % graphics
\usepackage{tikz}           % tikzpicture
\usepackage{quantikz}       % quantum circuits
\usepackage{algorithm} 
\usepackage{algorithmic}
\usepackage{xcolor}
\usepackage{capt-of}
\usepackage{verbatim}
\usepackage{caption}
\usepackage{subcaption}
\usepackage{rotating}
\usepackage{eqparbox}
\usepackage{float}
\usetikzlibrary{positioning}
\graphicspath{{media/}}     % organize your images and other figures under media/ folder
\newcommand{\sign}{\text{sign}}

\begin{document}

\preprint{APS/123-QED}

\title{\Large Variational Quantum Soft Actor-Critic for Robotic Arm Control}% Force line breaks with \\

\author{Alberto Acuto}
\email{alberto.acuto@nttdata.com}
\author{Paola Barillà}
\email{paola.barilla@nttdata.com}
\author{Ludovico Bozzolo}
\email{ludovico.bozzolo@nttdata.com}
\author{Matteo Conterno}
\email{matteo.conterno@nttdata.com}
\author{Mattia Pavese}
\email{mattia.pavese@nttdata.com}
\author{Antonio Policicchio}
\email{antonio.policicchio@nttdata.com}
\affiliation{NTT DATA Italia S.p.A.}
 %\altaffiliation{NTT DATA Italia S.p.A.}%Lines break automatically or can be forced with \\


\begin{abstract} 
%\begin{description}d
\begin{center}
\vspace{0.2cm}
\textbf{Abstract}
\end{center}
Deep Reinforcement Learning is emerging as a promising approach for the continuous control task of robotic arm movement. However, the challenges of learning robust and versatile control capabilities are still far from being resolved for real-world applications, mainly because of two common issues of this learning paradigm: the exploration strategy and the slow learning speed, sometimes known as ``the curse of dimensionality''. This work aims at exploring and assessing the advantages of the application of Quantum Computing to one of the state-of-art Reinforcement Learning techniques for continuous control -- namely Soft Actor-Critic. Specifically, the performance of a Variational Quantum Soft Actor-Critic on the movement of a virtual robotic arm has been investigated by means of digital simulations of quantum circuits. 
A quantum advantage over the classical algorithm has been found in terms of a significant decrease in the amount of required parameters for satisfactory model training, paving the way for further promising developments.
%\end{description}
\vspace{0.2cm}\\
\noindent\textit{\textbf{Keywords}}: Reinforcement Learning $\cdot$ Quantum Computing $\cdot$ Soft Actor-Critic $\cdot$ Robotic Arm
\end{abstract}

\maketitle

%\printglossary[type=\acronymtype]

%%%%%%%%%%%%%%%%%%%%%%%%%%%%%%%%%%%%%%%%%%%%%%%%%%%%%%%%%
%%%%%%%%%%%%%%%%%%%%%%%%%%%%%%%%%%%%%%%%%%%%%%%%%%%%%%%%%
%%%%%%%%%%%%%%%%%%%%%%%%%%%%%%%%%%%%%%%%%%%%%%%%%%%%%%%%%

\section{Introduction}
\label{sec:Introduction}
\input{01_Introduction.tex}

%%%%%%%%%%%%%%%%%%%%%%%%%%%%%%%%%%%%%%%%%%%%%%%%%%%%%%%%%
%%%%%%%%%%%%%%%%%%%%%%%%%%%%%%%%%%%%%%%%%%%%%%%%%%%%%%%%%
%%%%%%%%%%%%%%%%%%%%%%%%%%%%%%%%%%%%%%%%%%%%%%%%%%%%%%%%%

\section{Related Works}
\label{sec:soa}
\input{02_RelatedWorks.tex}

%%%%%%%%%%%%%%%%%%%%%%%%%%%%%%%%%%%%%%%%%%%%%%%%%%%%%%%%%
%%%%%%%%%%%%%%%%%%%%%%%%%%%%%%%%%%%%%%%%%%%%%%%%%%%%%%%%%
%%%%%%%%%%%%%%%%%%%%%%%%%%%%%%%%%%%%%%%%%%%%%%%%%%%%%%%%%

\section{Quantum Reinforcement Learning for robotic arm control}
\label{sec:QRL}
\input{03_QRL.tex}

%%%%%%%%%%%%%%%%%%%%%%%%%%%%%%%%%%%%%%%%%%%%%%%%%%%%%%%%%
%%%%%%%%%%%%%%%%%%%%%%%%%%%%%%%%%%%%%%%%%%%%%%%%%%%%%%%%%
%%%%%%%%%%%%%%%%%%%%%%%%%%%%%%%%%%%%%%%%%%%%%%%%%%%%%%%%%

\section{Robotic Arm Environment}
\label{sec:RAE}
\input{04_RoboticArmEnv.tex}

%%%%%%%%%%%%%%%%%%%%%%%%%%%%%%%%%%%%%%%%%%%%%%%%%%%%%%%%%
%%%%%%%%%%%%%%%%%%%%%%%%%%%%%%%%%%%%%%%%%%%%%%%%%%%%%%%%%
%%%%%%%%%%%%%%%%%%%%%%%%%%%%%%%%%%%%%%%%%%%%%%%%%%%%%%%%%

\section{Implementation of classical and Quantum Soft Actor-Critic algorithms}
\label{sec:exp_setup}
\input{05_AlgorithmImplementation.tex}

%%%%%%%%%%%%%%%%%%%%%%%%%%%%%%%%%%%%%%%%%%%%%%%%%%%%%%%%%
%%%%%%%%%%%%%%%%%%%%%%%%%%%%%%%%%%%%%%%%%%%%%%%%%%%%%%%%%
%%%%%%%%%%%%%%%%%%%%%%%%%%%%%%%%%%%%%%%%%%%%%%%%%%%%%%%%%
\section{Experiments}
\label{sec:exp}
\input{06_Experiments.tex}

%%%%%%%%%%%%%%%%%%%%%%%%%%%%%%%%%%%%%%%%%%%%%%%%%%%%%%%%%
%%%%%%%%%%%%%%%%%%%%%%%%%%%%%%%%%%%%%%%%%%%%%%%%%%%%%%%%%
%%%%%%%%%%%%%%%%%%%%%%%%%%%%%%%%%%%%%%%%%%%%%%%%%%%%%%%%%

\section{Conclusions and future works}
\label{sec:conclusions}
\input{07_Conclusions.tex}

%%%%%%%%%%%%%%%%%%%%%%%%%%%%%%%%%%%%%%%%%%%%%%%%%%%%%%%%%
%%%%%%%%%%%%%%%%%%%%%%%%%%%%%%%%%%%%%%%%%%%%%%%%%%%%%%%%%
%%%%%%%%%%%%%%%%%%%%%%%%%%%%%%%%%%%%%%%%%%%%%%%%%%%%%%%%%

\begin{acknowledgments}
The authors wish to acknowledge the support of Prof. Noah Klarmann of Rosenheim University of Applied Sciences, for providing the base Reinforcement Learning environment and for useful discussions.

\vfill
This work has been fully funded by NTT DATA Corporation.
\end{acknowledgments}

\newpage
\bibliography{references}

\end{document}

%% file: 01_Introduction.tex
Deep Reinforcement Learning (DRL) \cite{DRL}, the combination of Deep Learning and Reinforcement Learning (RL), has emerged in robotic control as a promising approach for autonomously acquiring complex behaviors from low-level sensor observations \cite{45926, Ibarz_2021, Liu_2021, Kilinc_2021}. Given the advances in DRL, and the developments in robotics and mechanics, the research community has searched for for more software-based control solutions for robots using low-cost sensors, having fewer requirements for the operating environment and calibration. The key is to focus on robust software algorithms: instead of hard-coding directions in Programmable Logic Controllers to govern the robot movement, the control policy could be obtained by learning, and then updated accordingly. Using RL-based techniques for robot control is appealing because it can enable robots to move towards unstructured environments, to handle unknown objects, and to learn a state representation suitable for multiple tasks. Compared to RL, DRL allows to fix critical issues relative to the dimensionality and scalability of data in tasks with sparse reward signals, such as robotic manipulation and control tasks. However, despite recent  improvements, the challenges of learning robust manipulation skills for robots with DRL are still far from being solved for real-world applications. This is mainly due to some well know issues with DRL: sample efficiency, generalization, and computing resources for training the learning algorithms \cite{Liu_2021}. 
Regarding DRL for robotic control, sample efficiency means how much data are needed to be collected in order to build an optimal policy to accomplish the designed task. According to \cite{Liu_2021}, several issues in robotics prevent an effective sample efficiency: the agent cannot receive a training set provided by the environment unilaterally, but rather information which is determined by both the actions it takes and the dynamics of the environment; although the agent aims at maximizing the long-term reward, it can only observe the immediate reward; there is no clear boundary between training and test phases, since the time the agent spends trying to improve the policy often comes at the expense of utilizing this policy, which is often referred to as the exploration–exploitation trade-off \cite{Sutton1998}. 
On the other hand, generalization refers to the capacity to use previous knowledge from a source environment to achieve a good performance in a target environment, and applicability for flexible long-term autonomy. This is widely seen as a necessary step to produce artificial intelligence that behaves similar to humans.  %One of the strategies followed to tackle with generalization is to train a model on a variety of learning tasks, such that it can solve new learning tasks using only a small number of training samples \cite{meta_learning}.
Moreover, it has to be noted that, given the large amount of data to reach optimal results, DRL is computationally intensive, and it requires high-performance computers for model training and fastening the learning process. More progress is required to overcome such limitations, as both gathering experiences by interacting with the environment and collecting expert demonstrations for RL are expensive procedures. 

Quantum Computing (QC) promises the availability of computational resources and generalization capabilities well beyond the possibilities of classical computers \cite{10.5555/1972505}. Even if fault-tolerant quantum devices are still far to come, near-term devices -- Noisy Intermediate-Scale Quantum Computers (NISQ) \cite{Preskill_2018}, limited in the number of qubits, coherence times and operations fidelity -- can already be utilized for a variety of problems. One promising approach is the hybrid training of Variational or Parameterized Quantum Circuits (VQCs or PQCs), i.e. the optimization of a parameterized quantum algorithm as a function approximation with classical optimization techniques \cite{BPQC}. VQCs are typically composed of fixed gates, e.g. controlled NOTs, and adjustable gates, e.g. qubit rotations parameterized by a set of free parameters. The main approach in the scientific community is to formalize problems of interest as variational optimization tasks, then using a hybrid quantum-classical hardware setup to find approximate solutions. By implementing some subroutines on classical hardware, the requirement of quantum resources is significantly reduced, particularly in the number of qubits, circuit depth, and coherence time. Therefore, in the hybrid algorithmic approach, NISQ hardware focuses entirely on the classically intractable part of the problem. 

Quantum Machine Learning (QML) typically involves training a VQC in order to analyze either classical or quantum data \cite{schuld2021machine}. QML models may offer some advantages over classical models in terms of memory consumption and sample complexity for classical data analysis. Moreover, a recent research presented a comprehensive study of generalization performance in QML after training on a limited number of training data points, showing that a good generalization is guaranteed from few training data \cite{Banchi_2021}. All such aspects look promising to overcome the DRL issues discussed above for robot control. 

Starting from recent utilization of VQCs in RL problems  \cite{https://doi.org/10.48550/arxiv.2202.12180, 9144562, articleQML}, the possible application of quantum-classical hybrid algorithms to the control task of a robotic arm has been investigated. Specifically, the advantages of the application of VQCs to one of the state-of-art Reinforcement Learning techniques for continuous control -- the Soft Actor-Critic (SAC) -- have been explored and assessed by means of digital simulations of quantum circuits. 
The work demonstrates that robot control is a viable field of application for Quantum Reinforcement Learning (QRL) and of future advancements in autonomous robotics. 

The paper is organized as follows. Section \ref{sec:soa} provides a brief overview of previous works on
DRL with VQCs. The QRL approach used in this work is presented in Section \ref{sec:QRL}. The robotic arm modeling the RL environment is described in Section \ref{sec:RAE}. Results of the implemented quantum algorithms compared to a classical benchmark are shown in Section \ref{sec:exp}.
Finally, conclusions and potential future developments are outlined in Section \ref{sec:conclusions}.

%% file: 02_RelatedWorks.tex
Although the application of Variational Quantum Circuits to Reinforcement Learning is a relatively new field of research, several interesting papers have already been published on this subject. This section reports on a selection of research works on QRL.
Chen et al. \cite{9144562} have explored the potential of VQC in DRL. In particular, the authors have reshaped classical DRL algorithms such as experience replay and target network, into a representation of VQCs. In this work they have demonstrated that VQCs can be used to approximate the deep $Q$-value function for decision making and policy-section RL with experience replay and target network.
Wang et al. \cite{pmlr-v139-wang21w} have designed a QRL algorithm that approximates an optimal policy, the optimal value function and the optimal $Q$-function, assuming the algorithms can access samples from the environment in quantum superposition. This quantum algorithm has achieved quadratic speedups over the classical counterpart.
Skolik et al. \cite{Skolik2021QuantumAI} have investigated architectural choices for the so called quantum $Q$-learning agents to solving environments taken from OpenAI Gym \cite{openai}, then showing how such quantum $Q$-learning algorithms depend on observables of the quantum model, and providing insights on how to choose them. 

In the robotics field, Liu et al. \cite{Liu_2021} have presented a significant progress in robotic control application of DRL, solving critical issues relative to the dimensionality and scalability of data in tasks with sparse reward signals, such as robotic manipulation and control tasks.
One of the state-of-the-art algorithms in classical RL for continuous control tasks is the Soft Actor-Critic, proposed by Haarnoja et al. in \cite{https://doi.org/10.48550/arxiv.1812.05905}. Learning stability and efficiency make this algorithm a promising candidate for learning in real-world robotics tasks. Lillicrap et al. \cite{https://doi.org/10.48550/arxiv.2112.11921} have proposed variational quantum version of the classical SAC in which only the Actor neural network was replaced by a VQC. Results using the Pendulum OpenAI Gym Environment \cite{openai} have shown a quantum advantage in reducing model parameters while achieving similar performance of the classical algorithm.

%% file: 03_QRL.tex
The task of robotic arm operation requires continuous control, due to the continuous-valued observations coming from sensors and actions for accurate movement control. In this section, Reinforcement Learning application to continuous control is discussed with particular emphasis to the Soft Actor-Critic approach. Then, basic notation of Quantum Computing is presented and the application of Variational Quantum Circuits to Reinforcement Learning is discussed.

%%%%%%%%%%%%%%%%%%%%%%%%%%%%%%%%%%%%%%%%%%%%%%%%%%%%%%%%%
%%%%%%%%%%%%%%%%%%%%%%%%%%%%%%%%%%%%%%%%%%%%%%%%%%%%%%%%%
\subsection{Reinforcement Learning for continuous control}
\label{sec:RL}
\begin{figure}[!ht]
\centering
\includegraphics[width=0.9\linewidth]{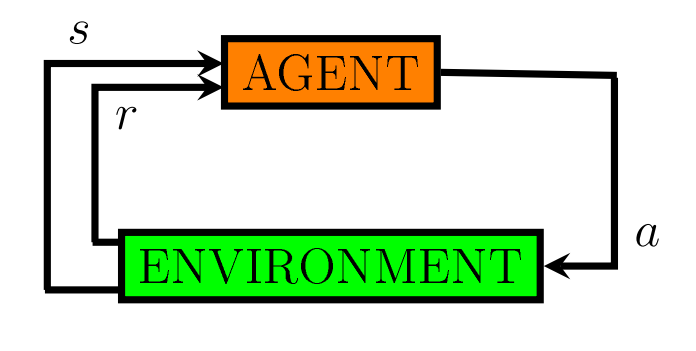}
\caption{Schematic model of Reinforcement Learning algorithm: $s$ is the environment state, $r$ is the reward, and $a$ is the agent action \cite{https://doi.org/10.48550/arxiv.1509.02971}.}
\label{fig:rl}
\end{figure}

Reinforcement Learning \cite{NIAN2020106887} is a Machine Learning technique in which an agent learns to behave in an environment by performing actions and assessing the results of those actions. For each good action, the agent gets positive reward, and for each bad action, the agent gets negative reward or penalty.
%consists of creating the most optimal agent capable to execute a predefined task through interaction with the environment by actions.
In a standard RL setup consisting of an agent interacting with an environment \textit{E} in discrete time steps, at each time step \textit{t} the agent receives an observation $x_{t}$, takes an action $a_{t}$, and receives a reward $r_{t}$.  A schematic representation is shown in Figure \ref{fig:rl}.\\
In general, the environment may be partially observed, so that the entire history of the observation and action pairs, $s_{t}= (x_{1}, a_{1}, ..., a_{t-1}, x_{t})$, is required to describe interaction environment-actor. Here, it is assumed the environment is fully observed, so that the environment state corresponds to the observation, i.e. $s_{t} = x_{t}$. The environment can be stochastic or deterministic and, in order to model it, a Markov decision process \cite{Puterman:1319893} is used with a state space $S$, an action space $A \in \mathbb{R}^{N}$, an initial state distribution $p(s_{1})$, a transition dynamics $p(s_{t+1}|s_{t}, a_{t})$, and a reward function $r(s_{t}, a_{t})$.
An agent's behavior is defined by a policy, $\pi$, which maps states to a probability distribution over the action space given the state space, $\pi: S \rightarrow P(A)$.
The return from the current state is defined as the sum of discounted future rewards, $R_{t} = \sum_{i=t}^{T} \gamma^{i-t}r(s_{i, a_{i}})$, with a discounting factor $\gamma \in [0,1]$. Note that the return depends on the action taken, therefore the policy $\pi$ affects it directly. The goal in RL is to maximize the return, and this is strictly correlated to learning the best policy able to reach this goal. A useful function to describe the expected return combined with the state $s_{t}$ and the action $a_{t}$ following the policy $\pi$, is the action-value $Q$-function which depends only on the environment \cite{https://doi.org/10.48550/arxiv.1509.02971}.
The action-value function is fundamental in RL, as it underlies one of the first learning algorithm: the $Q$-learning algorithm \cite{watkins1992q} and its extension to Deep $Q$-learning in the context of Deep Learning. $Q$-learning is the main algorithm of the Value-Based approach to RL. Value-Based approaches try to find or approximate the optimal value function, which is a mapping between an action and a value. The higher the value, the better the action.
The robotic control typically requires a continuous action space, and $Q$-learning cannot be applied due to the enormous number of possible configurations that need to be explored and the difficulty to converge to an optimal solution for the given task. Policy-Based algorithms have been proposed for continuous control \cite{https://doi.org/10.48550/arxiv.1509.02971}. Such algorithms try to find the optimal policy directly without leveraging the $Q$-value \cite{ac}. The Policy-Based approach has its most suitable applications in continuous and stochastic environments, where faster convergence is guaranteed. On the other hand, Value-Based approaches are more sample-efficient and steady \cite{valuepolicy}. 

%%%%%%%%%%%%%%%%%%%%%%%%%%%%%%%%%%%%%%%%%%%%%%%%%%%%%%%%%
%%%%%%%%%%%%%%%%%%%%%%%%%%%%%%%%%%%%%%%%%%%%%%%%%%%%%%%%%
\subsection{Soft Actor-Critic}
\label{sec:SAC}

The Actor-Critic DRL algorithm merges together Policy-Based and Valued-Based approaches \cite{ac}. The core idea is to split the model in two components: the Actor, that defines an action based on the state, and the Critic, that produces the state-values. The actor essentially controls how the agent behaves by learning the optimal policy, and the Critic evaluates the action by computing the value function. The training of the neural networks representing the Actor and the Critic is performed separately, using gradient descent \cite{actorcriticmethod}. As training proceeds, the Actor learns to produce better actions to reach the goal, and the Critic gets better at evaluating those actions. 
The training process of the model begins with a starting state of the environment being observed by the agent. Then the agent takes an action, leading the environment into a new state, and letting the agent achieve a reward (step). When a terminal state is encountered, a learning episode ends -- an episode being a collection of all subsequent steps. In this process the model optimization procedure is required, in order to update model parameters according to a specified loss function. For the Actor-Critic algorithm, the update of neural network weights occurs at every step, and not at the end of every episode. 
\begin{figure}[!ht]
\centering
\includegraphics[width=\linewidth]{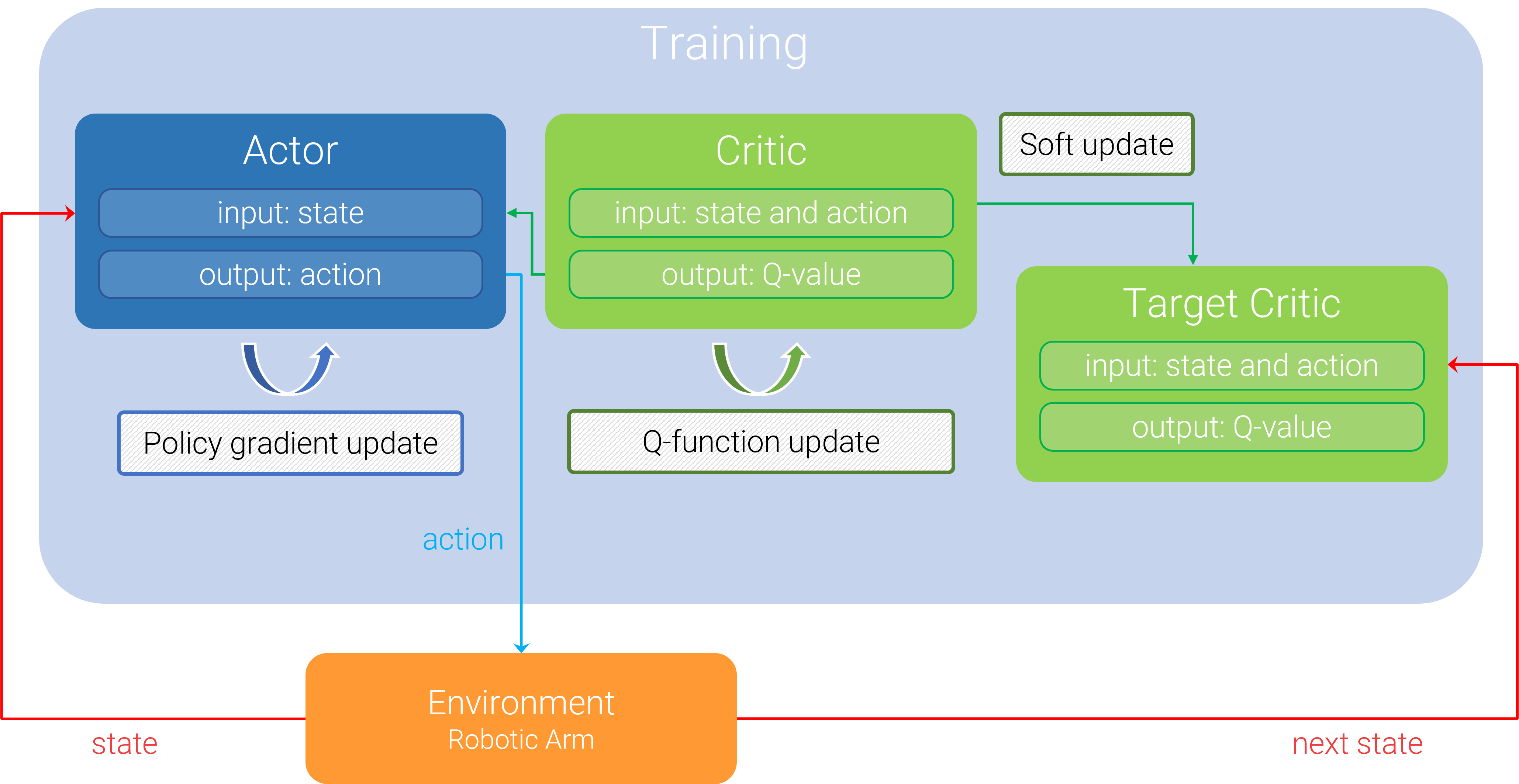}
\caption{Soft Actor-Critic training scheme.}
\label{fig:sac_training}
\end{figure}
The Actor-Critic approach suffers from high sensitivity to the values of hyperparameters. To tackle this problem and improve performance, stability and proficiency, the Soft Actor-Critic has been proposed by Haarnoja et al. in \cite{https://doi.org/10.48550/arxiv.1801.01290}. SAC introduces multiple elements to the Actor-Critic approach:
\begin{itemize}
    \item A large buffer memory to efficiently update the weights using the previous history, thus increasing learning stability.
    \item A target neural network of the Critic is introduced to further improve the stability and efficiency of learning the approximation of the value-state. To update the target Critic network, the soft function update procedure is used: values include a factor to combine new and old weights obtained from the updated and previous steps, and to this purpose a loss value of state and a loss of action must be defined.
    \item The maximum entropy of RL is introduced to increase the environment exploration, improve learning, and reduce sensitivity to hyperparameters. The loss of this maximum entropy was transformed for this case from the “classical” one considering the use of a neural network, and introducing an input noise vector sampled from a fixed spherical Gaussian distribution to calculate the action. In this context, maximum entropy represents the measure of chaos or disorder, strictly linked to the fact that the policy acts as randomly as possible to make the model more robust to variations.
\end{itemize}
Figure \ref{fig:sac_training} shows the model training process of a SAC algorithm.

%%%%%%%%%%%%%%%%%%%%%%%%%%%%%%%%%%%%%%%%%%%%%%%%%%%%%%%%%
%%%%%%%%%%%%%%%%%%%%%%%%%%%%%%%%%%%%%%%%%%%%%%%%%%%%%%%%%
\subsection{Quantum Computing}
\label{sec:QC}

Before describing the QRL approach to the SAC algorithm, a short introduction to common notations of Quantum Computing is given. For a comprehensive explanation of basic and advanced concepts we refer to general reviews such as  e.g., \cite{nielsen00}.

First off, it is worth describing the qubit beginning from its mathematical formulation. A quantum system evolves according to unitary evolution, defined as the time evolution of its quantum state following the Schrödinger equation and mathematically represented by a unitary operator. Unitary operators preserve the lengths and angles between vectors, and can be thought of as a kind of rotation operator in a vector space. When an experimental equipment observes the system to find out what is going on inside it, unitary evolution is interrupted by the external system. Such behavior is a consequence of the measurement process \cite{nielsen00}, defined by a set of measurement operators $\{M_m\}$, where the index \(m\) indicates possible measurement outcomes.

Classical bits can only be found in one of the two different states, 0 or 1. On the contrary, a qubit is a linear combination, or coherent superposition, of two states of the computational basis, called $|0\rangle$ and $|1\rangle$ with the bra-ket notation of Quantum Mechanics \cite{landau}. This notation refers to a mathematical representation of vectors in a complex vector space (Hilbert space), and a physical representation of vectors as states of a quantum system. The $|0\rangle$ and $|1\rangle$ states form an orthonormal basis into their vector space. In an orthonormal basis, vectors are all unit vectors, and orthogonal to each other. In QC, the computational basis is defined by tensors with following values: \[|0\rangle = \begin{bmatrix} 1 \\ 0 \end{bmatrix}, \ \ |1\rangle = \begin{bmatrix} 0 \\ 1 \end{bmatrix}.\] Hence, a qubit will be found in a general superposition state: \[|\psi\rangle = \alpha|0\rangle + \beta|1\rangle, \hspace{0.1cm} \text{with} \hspace{0.1cm} \alpha, \beta \in \mathbb{C}, \hspace{0.15cm} |\alpha|^2 + |\beta|^2 = 1.\] When a measurement is taken,  either 0 with $|\alpha|^2$ probability or 1 with $|\beta|^2$ probability can be  obtained. Before the measurement, the state can be modified by applying unitary operations as follows: \[|\psi'\rangle = U|\psi\rangle.\]
The above concepts are essential to introduce the so called gate model of quantum computation, implemented by quantum circuits. In Quantum Computing, unitary operators are represented by quantum gates such as \(X, Y, Z\) and CNOT gates, represented by square matrices. All this formalism can be extended for the case of multiple qubits, preparing quantum registers composed of \textit{n} qubits.
A quantum circuit is a group of unitary operations and measurements used to evolve a quantum state to a specific one.
To process data with Quantum Computers, classical data must be converted into quantum states. Quantum encoding techniques allow to do just that \cite{schuld2021machine}. Basically, an initial set of \(n\) states $|0\rangle^{\otimes n}$ is prepared -- the symbol "$\otimes$" representing the tensor product. For instance, considering two $|0\rangle$ states, the tensor product can be written as $|00\rangle$ or $|0\rangle|0\rangle$ or $|0\rangle \otimes |0\rangle$, and is defined as follows: \[|0\rangle^{\otimes 2} = |0\rangle \otimes |0\rangle = \begin{bmatrix} 1 \\ 0 \end{bmatrix} \otimes \begin{bmatrix} 1 \\ 0 \end{bmatrix} = \begin{bmatrix} 1 \begin{bmatrix} 1 \\ 0 \end{bmatrix} \\ 0 \begin{bmatrix} 1 \\ 0 \end{bmatrix} \end{bmatrix} = \begin{bmatrix} 1 \\ 0 \\ 0 \\ 0 \end{bmatrix}.\] Then, a set of unitary transformations are applied, in order to obtain the data encoding. In the present work, angle encoding has been exploited. Angle encoding makes use of rotation gates \(R\) to encode classical information \textbf{x}. Considering a system of \(n\) qubits, the classical information determines angles of rotation gates as follows: \[|\mathbf{x}\rangle = \bigotimes_i^nR(\mathbf{x}_i)|0^n\rangle.\]\\
In addition, parameterized gates have been used to define VQCs, such as:
\begin{equation*}
\begin{split}
    &R_x(\theta) = \exp\bigg(-i\frac{\theta}{2}X\bigg),
    %\qquad
    \\&R_y(\theta) = \exp\bigg(-i\frac{\theta}{2}Y\bigg),
    \\&R_z(\theta) = \exp\bigg(-i\frac{\theta}{2}Z\bigg).
\end{split}    
\end{equation*}
where $\theta$ is a variable controlling the gate.

%%%%%%%%%%%%%%%%%%%%%%%%%%%%%%%%%%%%%%%%%%%%%%%%%%%%%%%%%
%%%%%%%%%%%%%%%%%%%%%%%%%%%%%%%%%%%%%%%%%%%%%%%%%%%%%%%%%
\subsection{Variational Quantum Circuits for Reinforcement Learning}
\label{sec:VQC}
\begin{figure*}[!ht]
\centering
\includegraphics[width=0.5\textwidth]{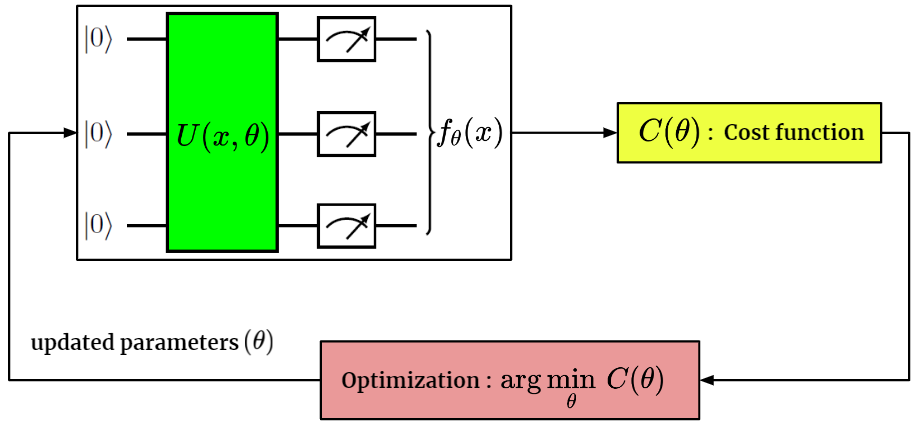}
\caption{Schematic representation of a Variational Quantum Algorithm.}
\label{fig:vqc}
\end{figure*}
A Variational Quantum Circuit consists of various quantum gates and wires, with learnable parameters to adjust these gates \cite{schuld2021machine}. Like the parameters in a neural network, these learnable parameters in VQCs can be optimized to approximate complex continuous functions. So, VQCs are also known as quantum neural networks. A typical VQC has three main components: encoder, variational circuit, and measure. The encoder is usually the first operation of any quantum circuits. Given a classical input $x$, the encoder encodes it into a quantum state.
%There are different types of encoder circuits: basis embedding, angle embedding and amplitude embedding. For our use case we choose the angle embedding that encodes \textit{n} binary features into a basis state of \textit{n} qubits. 
The variational circuit often follows an encoder circuit. Usually, it consists of alternate layers of rotation gates and entanglement gates (e.g., a closed chain or ring of CNOT gates). After the variational circuit, the states of qubits are measured.
Like with neural networks, a VQC is trained through gradient descent. Given a quantum simulator, backpropagation \cite{Rumelhart1986LearningRB} can be applied to compute gradient analytically and optimize the VQC parameters. However, backpropagation is not applicable for a physical Quantum Computer, since it is impossible to measure and store intermediate quantum states during computation without impacting the whole computation process. Instead, parameter-shift is used to compute gradients for any multi-qubit VQCs. This is done by evaluating the circuit when shifting the parameter by a specific quantity, and calculating the difference before and after the shifting \cite{https://doi.org/10.48550/arxiv.1905.13311}. After getting gradients, learnable parameters are optimized with traditional optimizers, such as RMSprop and Adam \cite{https://doi.org/10.48550/arxiv.2112.11921}. A schematic for this approach is shown in Figure \ref{fig:vqc}. Unlike neural networks, VQCs do not express non-linearity, except for measurement, unless a particularly encoding strategy is applied. In VQCs, it can be noticed that the same gate applied on all qubits can be seen as a neuron of a neural network layer \cite{Conterno}. Additionally, it can be observed how the measurement operations lead to non-linearity. To introduce more non-linearity, depolarizing gates could be added, but there is no theorem able to quantify how many of them are needed. This may be a problem in DRL, due to the significant presence of non-linear functions in the $Q$-values and Actor components. To introduce this non-linearity and deal with the no-cloning property of quantum computing, a new ansatz called \textit{data re-uploading} has been defined. The concept has been introduced in \cite{P_rez_Salinas_2020}, and is meant to introduce non-linearity by reapplying the encoding layer multiple times, to have a more composite expressiveness of the function. The strategy is therefore to use more quantum gates without increasing the number of required qubits. 

%% file: 04_RoboticArmEnv.tex
%%%%%%%%%%%%%%%%%%%%%%%%%%%%%%%%%%%%%%%%%%%%%%%%%%%%%
%%%%%%%%%%%%%%%%%%%%%%%%%%%%%%%%%%%%%%%%%%%%%%%%%%%%%
\subsection{Robotic arm modeling}
\begin{figure}[H]
\centering
\includegraphics[width=0.7\linewidth]{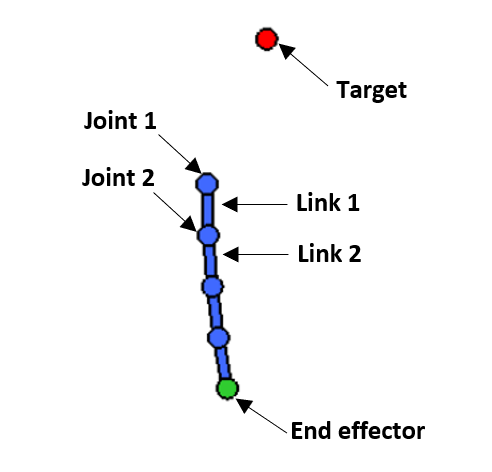}
\caption{Simplified schematic diagram of mechanical components of a four-joint robotic arm.}
\label{fig:env}
\end{figure}

A robotic arm can be described as a chain of links that are moved by joints containing motors to change the link position and orientation. Figure \ref{fig:env} shows the schematic diagram of a simple 2-dimensional, four-joints robotic arm mounted and fixed on the first joint. The arm is able to move the links using the joints on the 2-dimensional plane and can independently move each link clockwise and counter-clockwise up to a given velocity. Last joint is referred to as the end effector. 
Such environment has been created using Box2D \cite{box2d}, by adapting parts of OpenAI Gym environment Acrobot-v1 \cite{acrobot}. Support for continuous action and state spaces is enabled, and the dynamics of a planar manipulator with an arbitrary number of links can be simulated.
For the scope of this work, a two-link structure has been considered, with link 1 connecting the center (i.e joint 1) to the so called middle effector (i.e joint 2), and link 2 connecting the middle effector to the end effector. Physical parameters of the environment are summarized in Table \ref{table:1}. As with standard custom Gym environments, all parameters are customizable.
Let $\vec{x}_m = (x_m, y_m)$, $\vec{x}_e = (x_e, y_e)$ and $\vec{x}_t = (x_t, y_t)$ be the positions of the middle effector, end effector and target respectively. The environment allows for accessing the following states: 
target position $\vec{x}_t$; end effector position $\vec{x}_e$; rotation angles $(\theta, \phi)$ of the two links with respect to vertical axis.
Based on the state of the environment, an algorithm $\mathcal{A}$ defines the actions to perform, applying torques $(a_m, a_e)$ to the middle effector and to the end effector, with intensity ranging between $[- max\_torque$, $max\_torque]$.
Once an arbitrary number of episodes $N$ and a maximum number of steps per episode (250 in this case) are chosen, the logic flow shown in Algorithm \ref{alg:alg_env} begins. When an episode starts, the target position is randomly initialized, while the effector positions are always initialized so that the robotic arm is in the direction in which gravity pulls. Then, for each step in the episode, the agent gets state $s$ of the environment and computes, according to it and through the algorithm $\mathcal{A}$, the actions to take. After that, the agent computes the Euclidean distance $d$ between $\vec{x}_e$ and $\vec{x}_t$. Once a distance threshold $T$ is defined, reward $r$ for each step is set to $-d$ if $d>T$, otherwise it is set to $+5$; $r$ is then used to update the episode return $R$ (i.e, the sum of all rewards obtained in it). Episode terminates when $d \leq T$, or when maximum number of steps is reached. 

\begin{table*}[!ht]
\begin{center}
\begin{tabular}{ |c|c|c| } 
 \hline
 \footnotesize \bf Parameter & \footnotesize \bf Description & \footnotesize \bf Used Values \\ 
 \hline
 Link mass & Mass of links connecting the joints & 0.01 Kg \\ 
 \hline
 Link height & Vertical length of links & 0.5 m \\
 \hline
 Link width & Horizontal length of links & 0.1 m \\
 \hline
 Episode length & Maximum number of steps for every episode & 250 \\
 \hline
 Frames per second & Refresh rate for animation & 50 \\
 \hline
 Distance threshold & Distance for which the target is considered reached & 0.25 m\\
 \hline
 Max joint velocity & Maximum velocity of rotation of link & 2.5 rad/s\\
 \hline
 Max joint torque & Maximum torque that can be applied on joint & 1000 Nm\\
 \hline
\end{tabular}
%\vspace{0.3cm}
\caption{Two-links robotic arm environment parameters.}
\label{table:1}
\end{center}
\end{table*}%

\setlength{\textfloatsep}{0pt}
\begin{algorithm}[H]
    \begin{algorithmic}
    \REQUIRE number of episodes $N$
    \REQUIRE algorithm $\mathcal{A}$
    \FOR {$N$ episodes}
        \STATE{\textbf{init} target position $(x_t, y_t)$}
        \STATE{\textbf{init} effectors positions $(x_m, y_m), (x_e, y_e)$}
        \STATE{\textbf{init} episode reward $r = 0$}
        \FOR {250 steps}
            \STATE{\textbf{get} environment state $s$ }
            \STATE{\textbf{perform} actions $(a_m(\mathcal{A}, s), a_e(\mathcal{A}, s))$ } 
            \STATE{\textbf{get} new end effector position $(x_e, y_e)$}
            \STATE{\textbf{compute} $d \gets \sqrt{(x_t-x_e)^2 + (y_t-y_e)^2}$}
            \IF {$d > Threshold$}
                \STATE{$r \gets r - d$}
            \ELSE
                \STATE{$r \gets r + 5$}
                \STATE{\textbf{break}}
            \ENDIF
        \ENDFOR
    \ENDFOR
    \caption{Environment workflow}
    \label{alg:alg_env}
    \end{algorithmic}
\end{algorithm}
\vspace*{-0.2cm}

%%%%%%%%%%%%%%%%%%%%%%%%%%%%%%%%%%%%%%%%%%%%%%%%%%%%%%%%%
%%%%%%%%%%%%%%%%%%%%%%%%%%%%%%%%%%%%%%%%%%%%%%%%%%%%%%%%%
%%%%%%%%%%%%%%%%%%%%%%%%%%%%%%%%%%%%%%%%%%%%%%%%%%%%%%%%%
\subsection{Deterministic benchmark}
\label{sec:det_bench}
The structure of robotic arm environment considered for this work allows for precise resolution rules to be defined, regardless of the initial position of the target. \\
Assuming the two links are of the same length $L$, it is easy to deduce that, in general, there are two \textit{ideal configurations} allowing the end effector to reach the target, denoted by the tuples $(\theta^{*}_{1},\phi^{*}_{1} )$ and $(\theta^{*}_{2},\phi^{*}_{2} )$, where $\theta^{*}_{i}$ and $\phi^{*}_{i}$ respectively refer to the angles formed by link 1 and link 2, with respect to the vertical and growing clockwise. Formally, let $\vec{x}_c = (x_c,y_c)$ be the position of the center; while keeping valid the definitions of $\vec{x}_m$, $\vec{x}_e$ and $\vec{x}_t$ given previously, link 1 and 2 can be geometrically represented respectively by vectors $\vec{\ell}_{\theta} = \vec{x}_m - \vec{x}_c $ and $\vec{\ell}_{\phi} = \vec{x}_e - \vec{x}_m $. Solving the environment obviously requires $\vec{x}_e$ and $\vec{x}_t$ to match; assuming 
$\vec{d}_c = \vec{x}_t - \vec{x}_c $, both the ideal configurations presuppose $\vec{d}_c$, $\vec{\ell}_{\theta}$ and $\vec{\ell}_{\phi}$ to form an isosceles triangle with base $\vec{d}_c$. Letting $d_c$ be the length of $\vec{d}_c$, the angle $\alpha$ at the base of the isosceles triangle can be easily computed as $\alpha = \cos^{-1}(d_c/2L)$. From $\alpha$ and $\beta$, $\theta_{1}^{*}=(\beta - \alpha + 2\pi) \mod 2\pi$ and $\theta_{2}^{*}=(\beta + \alpha - 2\pi) \mod 2\pi$ can be computed. In order to compute $\phi_1^{*}$ and $\phi_2^{*}$, the ideal positions of the central effector are first obtained, knowing $\vec{x}_c$, $\theta_{i}^{*}$ and $L$; $\phi_{i}^{*}$ can then be retrieved with little effort, since $\vec{x}_t$ is known. Between $(\theta^{*}_{1},\phi^{*}_{1} ),(\theta^{*}_{2},\phi^{*}_{2} )$, the \textit{target configuration} $(\theta^*, \phi^*)$ is selected as the one that minimizes $\delta = \pi - ||\theta_{i}^* - \theta_{0}| - \pi|$, representing the angular distance between $\theta_i^*$ and $\theta_0$, with $\theta_0$ being the angle $\vec{\ell}_{\theta}$ forms with respect to the vertical when the environment is initialized (by default, $\theta_0 = 0$ radians).\\
Let $(\theta, \phi)$ be the the real angles formed by $\vec{\ell}_{\theta}$ and $\vec{\ell}_{\phi}$ respectively at each step; the proposed algorithm aims to minimize the distances $\Delta\theta = \theta^* - \theta$,  $\Delta\phi = \phi^* - \phi$, by sequentially applying torques $(a_{m}, a_{e})$ to the middle effector and to the end effector; a positive torque is applied counterclockwise, while a negative torque is applied clockwise. Reasoning about the physics of the problem, an efficient algorithm should take into account the advantage in exploiting the moment of inertia when both links move in the same direction; on the other hand, this argument does not hold for all ideal configurations as the target position varies: in some cases, it is convenient to move the two links in opposite directions. It is possible to distinguish two available set of actions, with the algorithm branching on the one to take depending on the position of the target and the resulting relationship between $\theta^*$ and $\phi^*$, as shown in Algorithm \ref{alg:det_bench}.
\begin{figure*}
     \centering
     \begin{subfigure}[h]{0.45\textwidth}
         \centering
         \includegraphics[width=\textwidth]{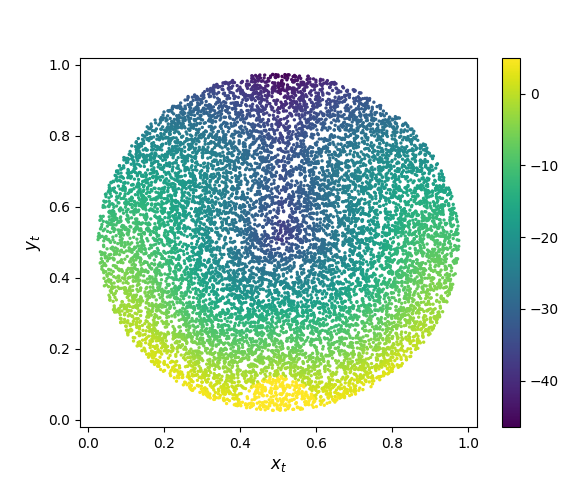}
         \caption{Returns distribution}
         \label{fig:return_distr}
     \end{subfigure}
     \hspace{0.5cm}
     \begin{subfigure}[h]{0.45\textwidth}
         %\centering
         \includegraphics[width=\textwidth]{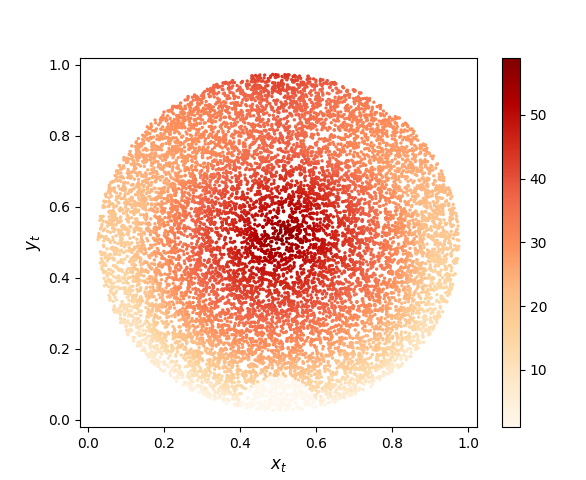}
         \caption{Step distribution}
         \label{fig:step_distr}
     \end{subfigure}
     \caption{Algorithm for deterministic benchmark: return and step distributions of the  two-links environment. $x_t$ and $y_t$ represent the positions of the target.} 
\label{fig:det_benchmark}
\end{figure*}
\begin{algorithm}[H]
\begin{algorithmic}
    \IF {$\theta^* < \pi$}
    \IF {$\theta^*<\phi^*<\theta^*+\pi$}
    \STATE {$(a_{m}, a_e) \gets (c_1 \sign(|\Delta\theta| - \pi)\Delta\theta, -c_2 \Delta\phi)$}
    \ELSE
    \STATE {$(a_{m}, a_e) \gets (c_1 \sign(|\Delta\theta| - \pi)\Delta\theta, c_2 \sign(|\Delta\phi| - \pi)\Delta\phi)$}
    \ENDIF
    \ELSIF {$\theta^* \geq \pi$}
    \IF {$\theta^*-\pi<\phi^*<\theta^*$}
    \STATE {$(a_{m}, a_e) \gets (c_1 \sign(|\Delta\theta| - \pi)\Delta\theta, -c_2 \Delta\phi)$}
    \ELSE
    \STATE {$(a_{m}, a_e) \gets (c_1 \sign(|\Delta\theta| - \pi)\Delta\theta, c_2 \sign(|\Delta\phi| - \pi)\Delta\phi)$}
    \ENDIF
    \ENDIF
\caption{Algorithm for deterministic benchmark}
\label{alg:det_bench}
\end{algorithmic}
\end{algorithm}
\begin{table}[H]
\begin{center}
    \begin{tabular}{ |c|c|c| } 
     \hline
     & \footnotesize \bf Steps & \footnotesize \bf Return \\
     \hline
     count & 8000 & 8000\\
     \hline
     mean & 30.252 & -17.397 \\
     \hline
     std & 11.347 & 11.528 \\
     \hline
     min & 1.000 & -47.055 \\
     \hline
     25\% & 23.000 & -26.691 \\
     \hline
     50\% & 31.000 & -17.431 \\
     \hline
     75\% & 38.000 & -8.149 \\
     \hline
     max & 58.000 & 5.000 \\
     \hline
    \end{tabular} 
    \vspace{0.3cm}
   \caption{Return and steps description of the algorithm for deterministic benchmark.}
   \label{table:2}
    \end{center}
\end{table}
Here, $c_1$ and $c_2$ are two real constants optimized trough a grid search. Whenever possible, the set of actions $(c_1 \sign(|\Delta\theta| - \pi)\Delta\theta, -c_2 \Delta\phi)$ allows for exploitation of the moment of inertia; both this set and the set $(c_1 \sign(|\Delta\theta| - \pi)\Delta\theta, c_2 \sign(|\Delta\phi| - \pi)\Delta\phi)$ ensure that the end effector will reach the target, regardless of the position of the latter. As the presented algorithm is always successful in solving the environment task in a reasonably small amount of steps, it can be considered as the reference \textit{optimal policy}, to which comparing the performance of the investigated RL and QRL algorithms. Performances of deterministic benchmark are shown in and Table \ref{table:2} and Figure \ref{fig:det_benchmark}.In the return and step distributions shown in the figure, the regions in the bottom part with the maximum return, represent the points for which the robotic arm requires only one step to reach the target, and there is no negative step reward.

%% file: 05_AlgorithmImplementation.tex
%%%%%%%%%%%%%%%%%%%%%%%%%%%%%%%%%%%%%%%%%%%%%%%%
%%%%%%%%%%%%%%%%%%%%%%%%%%%%%%%%%%%%%%%%%%%%%%%%
\subsection{Soft Actor-Critic algorithm}
\label{sec:SAC_IMPLEMENTATION}
\begin{figure*}
     \centering
     \begin{subfigure}[h]{0.74\textwidth}
         \centering
         \includegraphics[width=\textwidth]{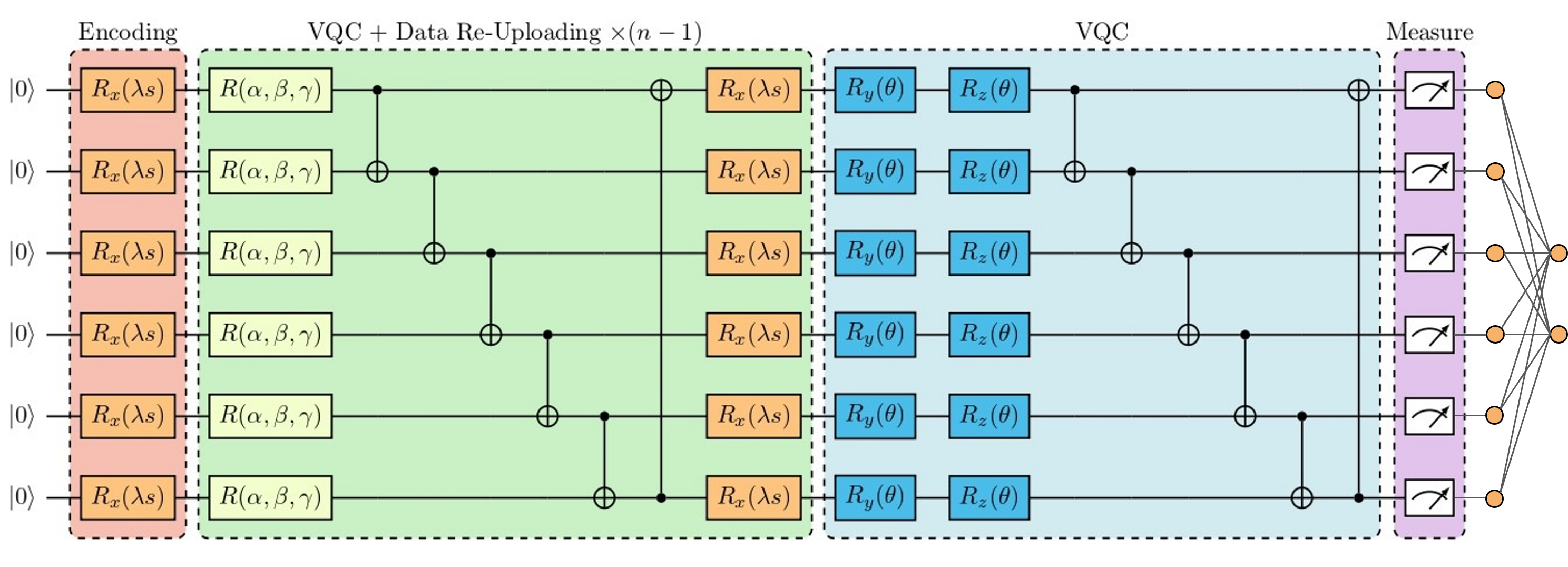}
         \caption{The architecture of quantum-classical hybrid Actor component of the QSAC}
         \label{fig:actorstructure}
     \end{subfigure}
     \begin{subfigure}[h]{0.83\textwidth}
         \centering
         \includegraphics[width=\textwidth]{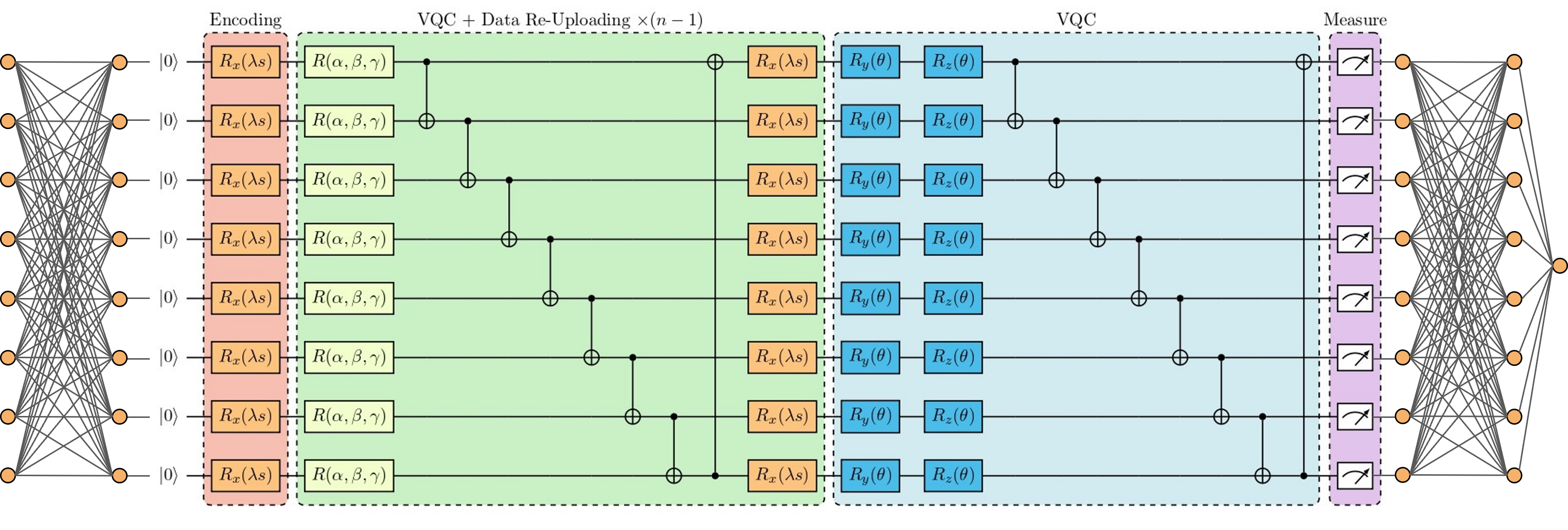}
         \caption{The architecture of quantum-classical hybrid Critic and target Critic components of the QSAC}
         \label{fig:criticstructure}
     \end{subfigure}
     \caption{The architecture of quantum-classical hybrid Actor, Critic and target Critic components of the QSAC.}
\label{fig:vqcstructure}
\end{figure*}
Soft Actor-Critic has been proposed in \cite{https://doi.org/10.48550/arxiv.1801.01290}. Its implementation replicates the algorithm structure as proposed in \cite{https://doi.org/10.48550/arxiv.2112.11921} and is summarized in the pseudo-code shown in Algorithm \ref{algo:SAC}. Indeed, the present work has started from the quantum-classical hybrid RL model proposed in \cite{https://doi.org/10.48550/arxiv.2112.11921}, to solve the Pendulum-v0 OpenAIGym  environment \cite{openai}.
The main components of the Soft Actor-Critic algorithm are: one Actor, two Critics and two target Critics. The Critic and target Critic networks share the same architecture. Due to the deep learning nature of this algorithm, all of these components are neural networks. Inspired by the Deep $Q$-Learning approach, there are target Critic components since the data need to be independent and identically distributed (``i.i.d.'') to correctly model the Q-functions: unfortunately, any two consecutive states do not have this property, and an independent copy of the Critic - the target - is necessary to meet this condition.
Moreover, recent advances have shown that using a copy for both Critic and target Critic components improves the performance and the stability of the algorithm.
The Critic networks are used to calculate the $Q$-functions of the actual state, while the target Critics are used to calculate the $Q$-functions of the next state, by choosing the minimum of the target outputs. These functions, as can be seen in Algorithm \ref{algo:SAC}, are used for optimization of the Critic components. 
The Actor components outputs two values for this algorithm: the mean and the variance of a Gaussian distribution that will be used to sample the actions used for each state. This means that the approach is statistical, but it can be converted to deterministic by using the mean of the distribution.
\begin{algorithm}[H]     
	\begin{algorithmic}
		\REQUIRE initial actor parameters $\theta$, initial critics parameters $\phi_1$ and $\phi_2$, $\gamma$, $\alpha$, $\rho$, empty experience replay $D$.
		\STATE Initialize the actor network with $\theta$.
		\STATE Initialize two critics networks with $\phi_1$ and $\phi_2$ respectively.
		\STATE Set target critics: $\phi_{targ, 1} \leftarrow \phi_1$ and $\phi_{targ, 2} \leftarrow \phi_1$.
		\FOR{each time-step}
		\STATE Observe state $S$, select action $A \sim \pi_{\theta}(\cdot |S)$ and execute $A$ in the environment.
		\STATE Observe next state $S'$, reward $R$, and binary done signal $d$ to indicate whether $S'$ is terminal state or not.
		\STATE Store $(S,A,R,S',d)$ in $D$.
		\STATE Reset the environment if $d=1$.
		\STATE Sample a batch of transitions $B = {(S,A,R,S',d)}$ from $D$ randomly.
		\STATE Compute target values $y(R,S',d) = R + \gamma (1-d) (min_{i=1,2}Q_{\phi_{targ,i}} (S',A') - \alpha \log \pi_{\theta}(A',S'))$ where $A' \sim \pi(\cdot,S')$.
		\STATE Update $\phi_i$ by minimizing: $\mathbb{E}_B [(Q_{\phi_i}(S,A) - y(R,S',d))^2]$ for $i = 1,2$.
		\STATE Update $\theta$ by maximizing: $\mathbb{E}_B [min_{i=1,2} Q_{\phi_i}(S,\tilde{A}_\theta) - \alpha \log \pi_{\theta}(\tilde{A}_\theta|S)]$.
		\STATE Do a soft update for target action-value networks: $\phi_{targ,i} \leftarrow \rho \phi_{targ,i} + (1-\rho)\phi_i$ for $i=1,2$.
		\ENDFOR
	    \caption{Soft-Actor-Critic algorithm}
        \label{algo:SAC}
	\end{algorithmic}
\end{algorithm}

%%%%%%%%%%%%%%%%%%%%%%%%%%%%%%%%%%%%%%%%%%%%%%%%
%%%%%%%%%%%%%%%%%%%%%%%%%%%%%%%%%%%%%%%%%%%%%%%%
\subsection{Variational Quantum Soft Actor-Critic algorithm}
\label{sec:QSAC}

The Quantum Soft Actor-Critic algorithm is quite similar to its classical counterpart, the only difference being the replacement of some neural network layers with a VQC. There is a major issue when using VQCs, namely the difficulty to approximate non-linear functions. In order to address it, two solutions already tested in \cite{https://doi.org/10.48550/arxiv.2112.11921} have been used: 
\begin{itemize}
    \item data re-uploading: a technique introduced in \cite{P_rez_Salinas_2020} where every VQC layer, except the last one, reapplies data encoding to have a more composite function expressiveness.
    \item neural networks: layers of neural networks are added before (for quantum Critic only) and after (for quantum Actor and Critic components) to leverage the high capacity of non-linearity using activation functions.
\end{itemize}

The first implementation explored in this work is identical to that proposed in \cite{https://doi.org/10.48550/arxiv.2112.11921}, where only the Actor neural network has been modified. The used structure is shown in Figure \ref{fig:actorstructure}. The VQC replaces the entire input layer in the hybrid structure, and neural network layers are later used. This allows for dimensional flexibility and the capacity to exploit potential quantum advantages. The encoding layer of the VQC implements the angle encoding technique, and every gate is parametrized to increase capacity to find a useful representation. The ansatz, that is repeated $n-1$ times, is composed by: a generic rotation over the three axis controlled by three different parameters, the CNOT gates to introduce entanglement, and a further encoding layer using data re-uploading. The last ansatz is composed by only $y$ and $z$-axis rotations and CNOT gates. Finally, measurement is applied using the Pauli-$Z$ gate.\\
In  a second implementation, the Critic neural network has been modified replacing some hidden layers with a VQC, while the Actor is a classical neural network. Figure \ref{fig:criticstructure} shows the structure of the hybrid circuit of the Critic component. As it can be seen the VQC, apart from the number of qubits, is equal to the one used for the actor. The real difference is the presence of multiple neural network layers before and after the VQC.\\
In a third QSAC implementation, denoted as Full QSAC, all Actor, Critic and target Critic components leverage the quantum-classical hybrid components shown in Figures \ref{fig:actorstructure} and  \ref{fig:criticstructure}.

%% file: 06_Experiments.tex
% \textcolor{red}{Citare tutte le tabelle nel testo. In generale essere più rigorosi, scientifici e quantitativi nel testo. Uniformare lo stile delle figure.}

%%%%%%%%%%%%%%%%%%%%%%%%%%%%%%%%%%%%%%%%%%%%%%%%%%%%%%%%%
%%%%%%%%%%%%%%%%%%%%%%%%%%%%%%%%%%%%%%%%%%%%%%%%%%%%%%%%%
\subsection{Simulation setup}
\label{sec:sim_set}
\begin{figure*}
     \centering
     \begin{subfigure}[h]{0.48\textwidth}
         \centering
         \includegraphics[width=\textwidth]{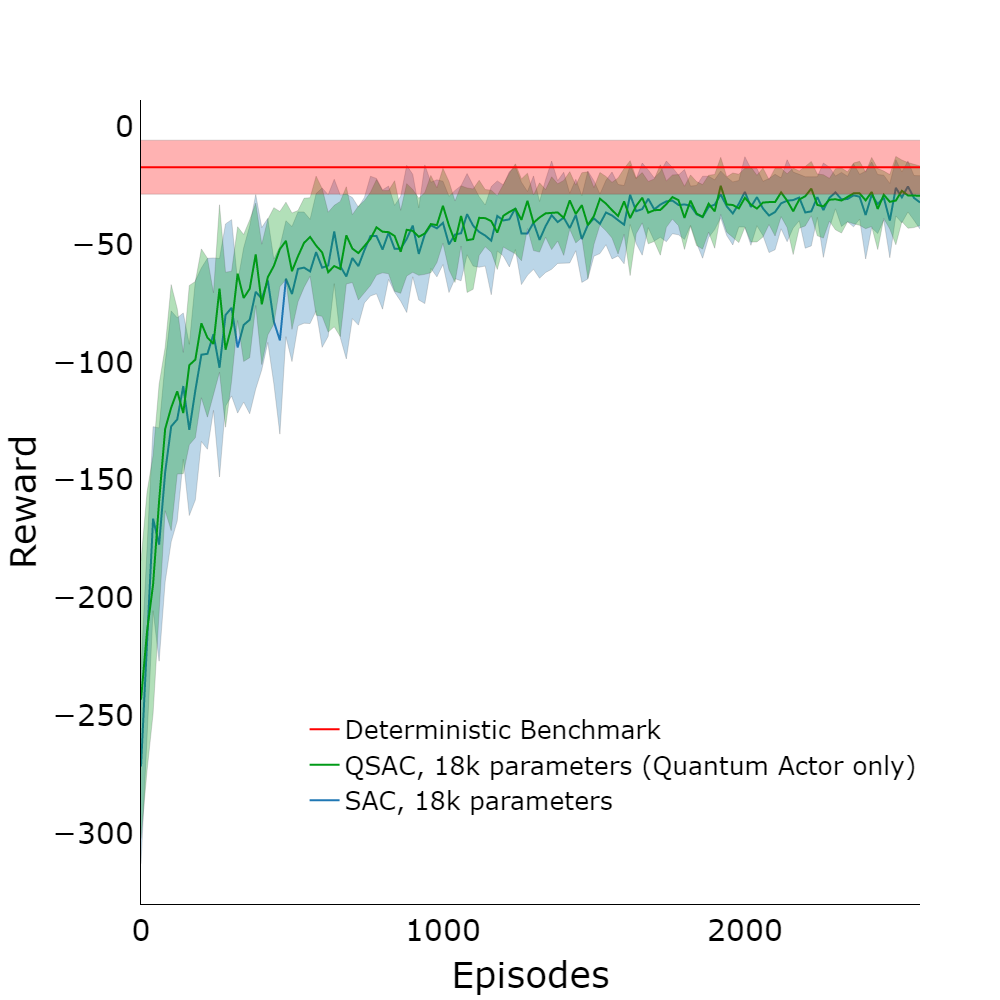}
         \caption{QSAC with hybrid Actor component only}
         \label{fig:qsac_actorquantum}
     \end{subfigure}
     \begin{subfigure}[h]{0.48\textwidth}
         %\centering
         \includegraphics[width=\textwidth]{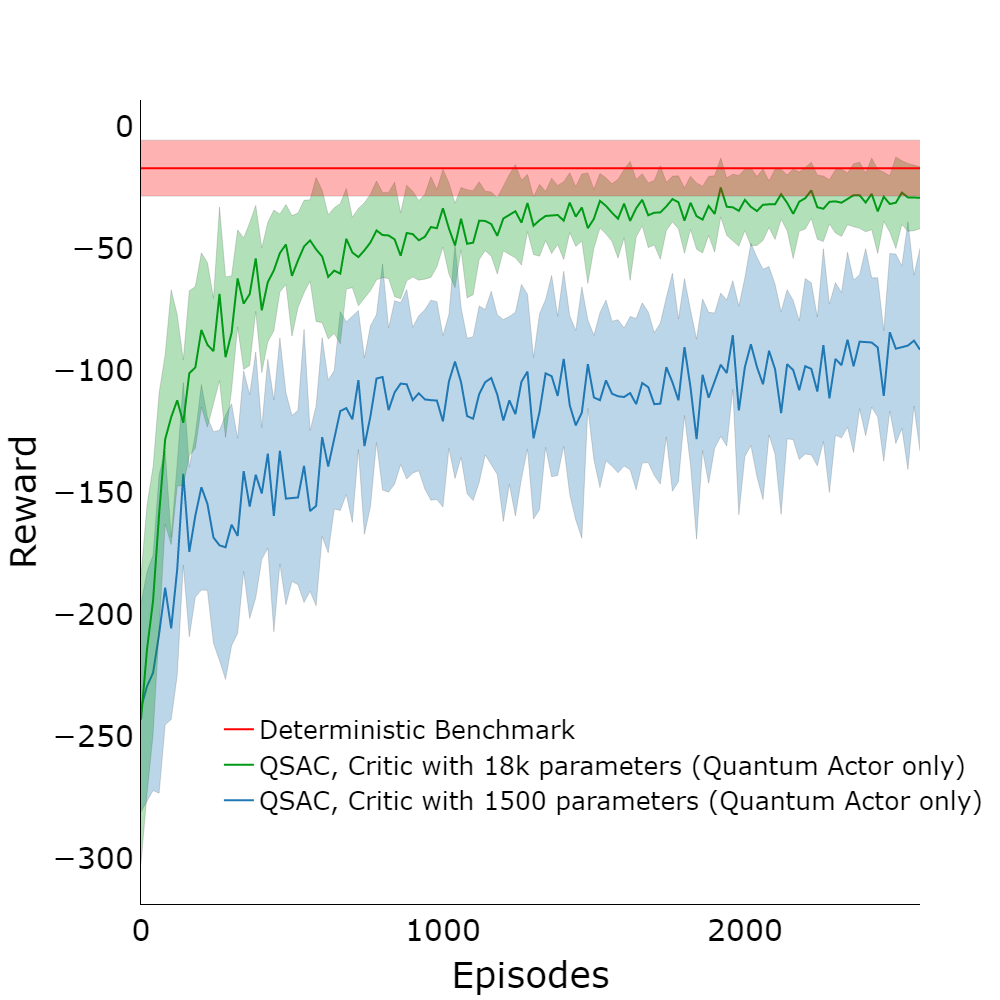}
         \caption{QSAC with hybrid Actor only and reduced Critic}
         \label{fig:qsac_criticreduced}
     \end{subfigure}
      \begin{subfigure}[h]{0.48\textwidth}
         %\centering
         \includegraphics[width=\textwidth]{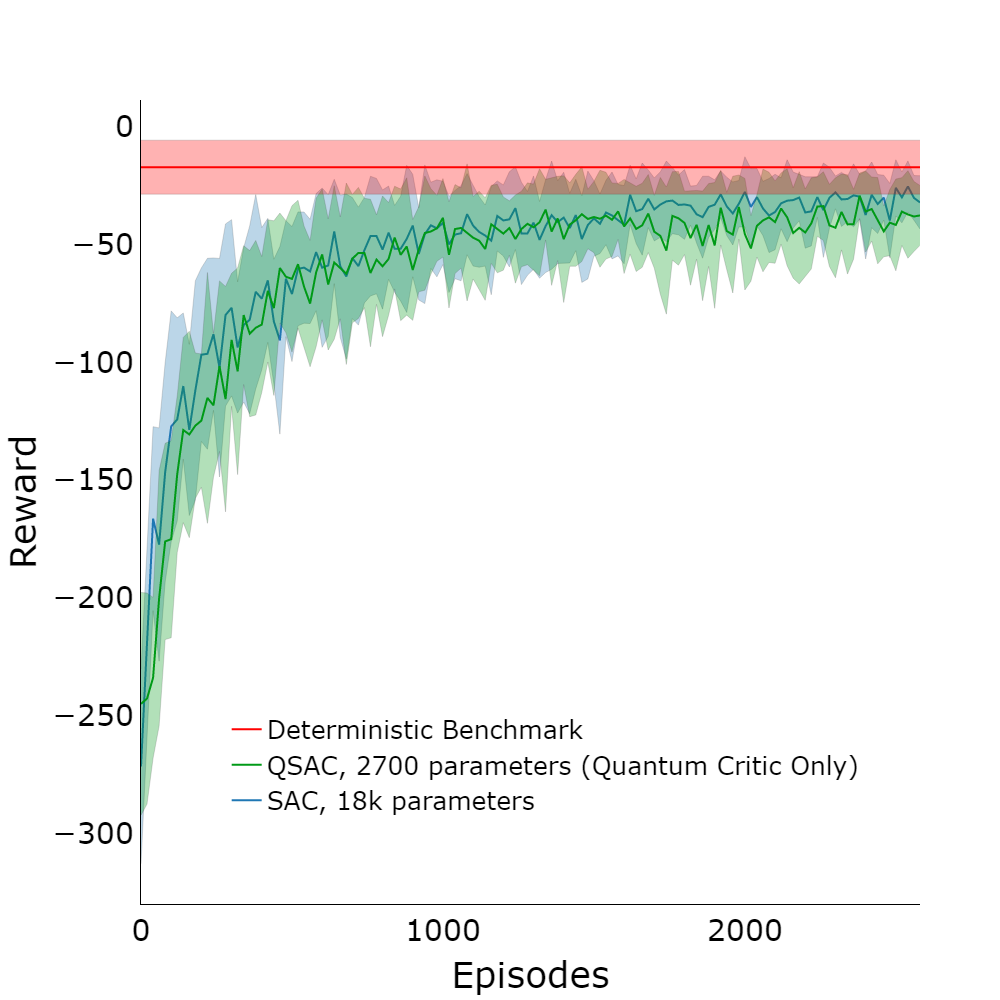}
         \caption{QSAC with hybrid Critic  components}
         \label{fig:qsac_criticquantum}
     \end{subfigure}      
     \begin{subfigure}[h]{0.48\textwidth}
         %\centering
         \includegraphics[width=\textwidth]{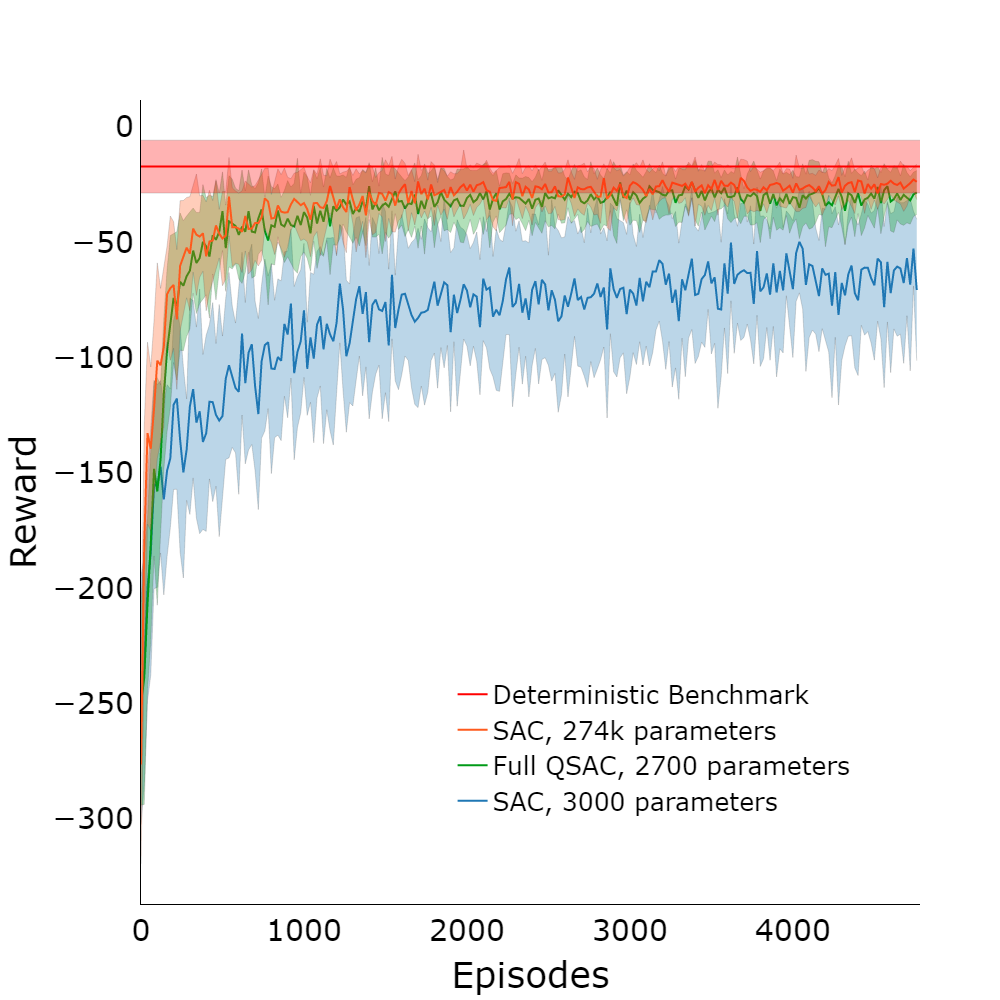}
         \caption{Full QSAC}
         \label{fig:fullqsac}
     \end{subfigure}
     \caption{Learning curves of classical and quantum-classical SAC architectures tested on the robotic arm environment.}
\label{fig:hybridplot}
\end{figure*}
Both classical and quantum SAC algorithms have been implemented using the Python programming language \cite{python}. Also the robotic arm environment has been developed in the same language. 
TensorFlow Quantum \cite{https://doi.org/10.48550/arxiv.2003.02989} has been used as the development framework for Quantum Machine Learning. This library works by simulating the quantum circuit using Cirq \cite{cirq_developers_2022_6599601} and distributing the computational workload through the multi-threading features of TensorFlow \cite{tensorflow2015-whitepaper}.
All classical components of the quantum-classical hybrid algorithm have been implemented using the TensorFlow library, in order to improve performance, reduce memory consumption, reduce dependencies, and avoid possible conflicts and errors during execution. Due to the simulation nature and the long time required to train this model, it was chosen not to explore how the introduction of noise may affect this kind of algorithm.
The simulation and training processes have been executed on a Virtual Machine equipped with 16 GB RAM, Intel Xeon CPU and Ubuntu 22.04 Operating System. %The training has used only CPU because the computational complexity has been particularly low due to the small circuit size. The other reason why no GPU has been used is that currently no Python libraries support GPU simulation stably.

The QSAC architectures presented in Section \ref{sec:SAC_IMPLEMENTATION} have been tested against several configurations of the neural networks and the VQC learnable parameters and hyperparameters.  %Two kind of quantum-hybrid SAC architectures have been tested: a quantum SAC with hybrid quantum-classical actor and a classical critic neural network, and a quantum SAC with both actor and critic hybrid quantum-classical. 
Each configuration has been compared to a classical SAC implementation of equivalent complexity in terms of the learnable parameters, using whenever possible the same exact hyperparameter values, optimizer and activation functions.
The environment is marked as ``solved'' according to two criteria: the first is relative to the deterministic benchmark described in Subsection \ref{sec:det_bench}, while the second is considered in absolute terms with respect to the RL and QRL algorithms considered. About the former, it has been required that the mean of returns for the last 1000 episodes of training falls within the range $[\mu - \sigma, \mu]$, where $\mu$ is the average mean of the return distribution of the deterministic benchmark, and $\sigma$ its standard deviation. Regarding the latter, it has been required for a maximum of 1\% of episodes to fail in the last 1000 episodes of the training. In particular, an episode is considered failed when the end effector does not reach the target within 250 steps. Both criteria have to be satisfied before 5000 episodes are reached. \\
All presented plots in the next sub-sections refer to an average of 20 episodes over 10 runs, with the shaded area representing one standard deviation. All networks have been optimized through Adam \cite{adam}, and the chosen hyperparameters have been taken from \cite{https://doi.org/10.48550/arxiv.2112.11921}. During the training process, several hyperparameter configurations have been tested without significant variations in the model performance.
%%%%%%%%%%%%%%%%%%%%%%%%%%%%%%%%%%%%%%%%%%%%%%%%%%%%%%%%%%
%%%%%%%%%%%%%%%%%%%%%%%%%%%%%%%%%%%%%%%%%%%%%%%%%%%%%%%%%%

\subsection{Quantum Soft Actor-Critic with hybrid quantum-classical Actor}
\label{sec:hybrid_actor}
First tests have been performed on the hybrid quantum-classical architecture only applied to the Actor component of the SAC. The Actor component must reproduce the mean and standard deviation of the normal distribution of actions.
The convergence criteria described in Subsection \ref{sec:sim_set} are satisfied  by the hybrid architecture shown in Figure \ref{fig:actorstructure} using four quantum layers and two classical layers, resulting in a total of 100 learnable parameters.
The Critics and target Critics are classical neural networks consisting of two hidden layers with 64 neurons, for a total of 18000 learnable parameters.
This hybrid structure has been compared with a classical neural network composed of a similar number of parameters for the Actor component, and the same structure for the Critic components. Results obtained using both hybrid and classical components are shown in Figure \ref{fig:qsac_actorquantum} together with the deterministic benchmark. The two architectures, QSAC represented in green and SAC represented in blue, share the same learning curve. No significant quantum advantage is observed in the ability to learn.\\
More configurations have been tested: in one case, the number of parameters of the Actor component in SAC and QSAC has been doubled, while leaving the Critic components unchanged, but the ability to converge was not influenced by that. 
Afterwards, another test has been conducted where the Critic components were reduced and the hybrid quantum Actor component was left unchanged. As shown in Figure \ref{fig:qsac_criticreduced} depicted in blue, reducing the size of the critic network components (each one reduced by ten times, as it can be seen in Table \ref{table:configurations1}) has made the algorithm unable to converge. 
It is thus possible to argue that the most impacting component on the ability to reach convergence of this algorithm is not the Actor, but the Critic. This may be due to the higher capacity of big neural networks to better approximate $Q$-functions compared to the small ones.
%%%%%%%%%%%%%%%%%%%%%%%%%%%%%%%%%%%%%%%%%%%%%%%%%%%%%%%%%%
%%%%%%%%%%%%%%%%%%%%%%%%%%%%%%%%%%%%%%%%%%%%%%%%%%%%%%%%%%
\subsection{Quantum Soft Actor-Critic with hybrid quantum-classical Critic}
\label{sec:hybrid_critic}
Taking into account the previous results obtained by reducing the critic components, a further test has been conducted. In this instance, configuration of the algorithm has been changed: the Actor has been kept classical, while the Critic components use the hybrid structure shown in Figure \ref{fig:criticstructure}. 
The Actor neural network consists of two classical hidden layers of 7 and 8 neurons respectively for a total of 149 learnable parameters, while the Critic and target Critic VQCs have 20 parametric layers and two hidden classical layers of 6 neurons each, for a total of 2600 learnable parameters.
Table \ref{table:configurations1} summarizes the configurations of the QSAC architectures. Results in Figure \ref{fig:qsac_criticquantum}, where the new configuration is depicted in green, show the same curve as in the classical case (curve in blue), but with an amount of parameters more than 6 times lower.
%%%%%%%%%%%%%%%%%%%%%%%%%%%%%%%%%%%%%%%%%%%%%%%%%%%%%%%%%%
%%%%%%%%%%%%%%%%%%%%%%%%%%%%%%%%%%%%%%%%%%%%%%%%%%%%%%%%%%
\subsection{Full Quantum Soft Actor-Critic}
\label{sub:full_hybrid}
\begin{figure}[H]
\centering
\includegraphics[width=0.95\linewidth]
{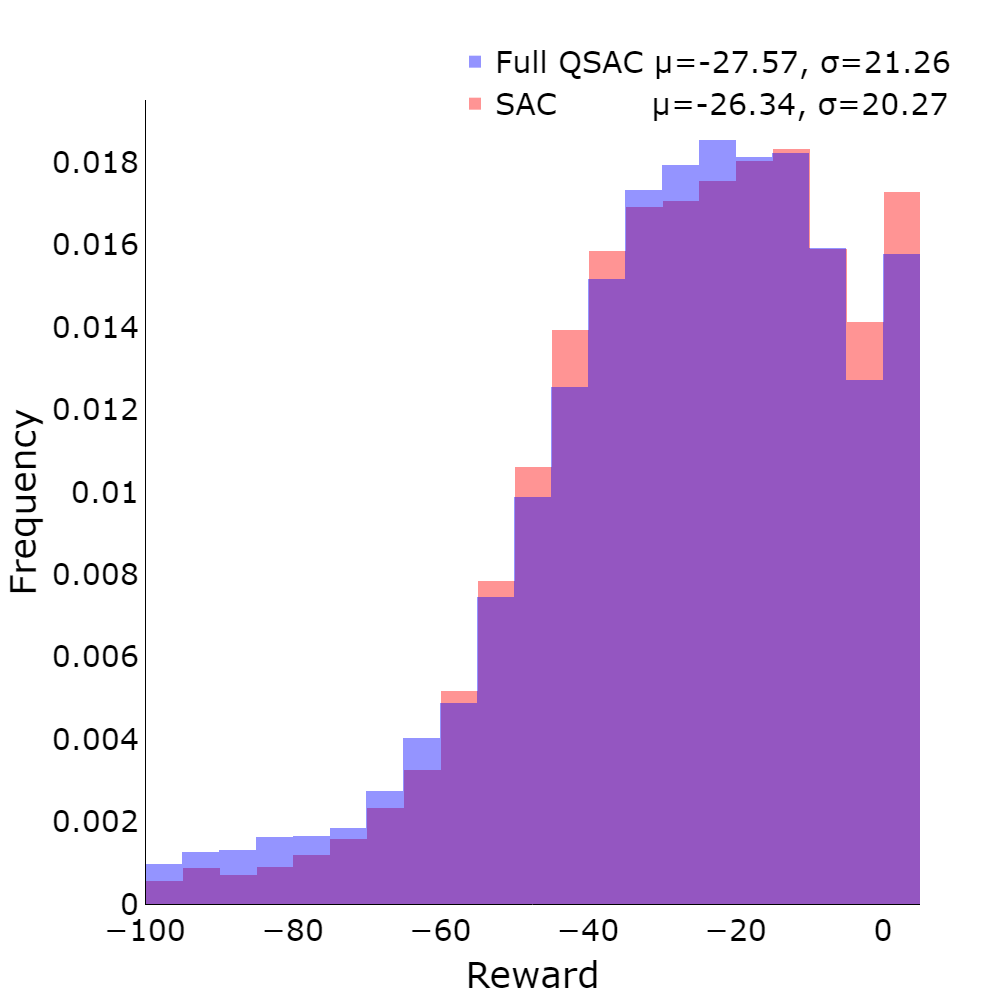}
\caption{Return  distribution of Full QSAC and SAC.}
\label{fig:qsac_distr}
\end{figure}

Results obtained by only tuning the Critic or the Actor component have given valuable insights to better understand where to expect potential quantum advantages. In this last configuration, all components are replaced by the quantum-classical hybrid architectures that have been previously described and shown in Figure \ref{fig:vqcstructure}, to understand what kind of advantages can be further obtained. Figure \ref{fig:fullqsac} shows a comparison of three training curves and the deterministic benchmark: in red representing deterministic benchmark, green Full QSAC with 2700 total learnable parameters, blue SAC with 3000 parameters (i.e. about the same amount of the QSAC) and orange SAC with 100 times the number of parameters of the QSAC (i.e. 274k parameters). A clear quantum advantage has been found in the number of learnable parameters. In particular, the classical SAC with the same number of parameters as the Full QSAC does not converge. The SAC requires 100 times the amount of parameters compared to the quantum algorithm to solve the environment with the same learning curve. Other checks have been done to verify if this QSAC algorithm presents more interesting properties. 
Indeed, Figure \ref{fig:qsac_distr} contains a comparison between the reward frequency distribution of the SAC (red histogram) and the Full QSAC (blue histogram), where advantages in terms of model stability cannot be noticed. 

\begin{table*}
\begin{center}
\begin{tabular}{|l|c|c|c|c|c|}\hline
\diagbox[width = 2.6 cm, height = 1.4 cm]{\footnotesize \bf Hyperparams}{\bf \footnotesize Model}
&{\footnotesize \bf SAC}&{\begin{minipage}{2.1 cm}\footnotesize \bf QSAC quantum actor only \end{minipage}}&{\begin{minipage}{2.1 cm}\footnotesize \bf QSAC quantum actor only, critic reduced by a factor 10 \end{minipage}}
&{\begin{minipage}{2.1cm} \footnotesize \bf QSAC \\ quantum critic only \end{minipage}}\\\hline
\hfil \footnotesize $\gamma$ & \footnotesize 0.99 & \footnotesize 0.99 & \footnotesize 0.99 & \footnotesize 0.99\\\hline
\hfil \footnotesize $\alpha$ & \footnotesize 0.2 & \footnotesize 0.2 & \footnotesize 0.2 & \footnotesize 0.2\\\hline
\hfil \footnotesize Learning rate & \footnotesize 0.0003 & \footnotesize 0.0003 & \footnotesize 0.0003 & \footnotesize 0.0003\\\hline
\hfil \footnotesize Memory size & \footnotesize 1000000 & \footnotesize 1000000 & \footnotesize 1000000 & \footnotesize 1000000\\\hline
\hfil \footnotesize Optimizer & \footnotesize Adam & \footnotesize Adam & \footnotesize Adam & \footnotesize Adam\\\hline
\hfil \footnotesize Actor neurons & \footnotesize (6,7)(8,(1,1)) & \footnotesize (6,VQA(4 layers),(1,1)) & \footnotesize (6, VQA(4 layers), (1,1)) & \footnotesize (6,7)(8,(1,1)) \footnotesize\\\hline
\hfil \footnotesize Actor act. func & \footnotesize (linear,relu,linear) & \footnotesize (linear,relu,linear) & \footnotesize (linear,relu,linear) & \footnotesize (linear,relu,linear)\\\hline
\hfil \footnotesize Actor params & \footnotesize 149 & \footnotesize 100 & \footnotesize 100 & \footnotesize 149\\\hline
\hfil \footnotesize Critic neurons & \footnotesize (8,64,64,1) & \footnotesize (8,64)(64,64)(64,1) & \footnotesize (8,16)(16,16)(16,1) & \footnotesize (8,VQA(20 layers),8,1)\\\hline
\hfil \footnotesize Critic act. func. & \footnotesize (linear,relu,relu,linear) & \footnotesize (linear,relu,relu,linear) & \footnotesize (linear,relu,relu,linear) & \footnotesize (linear,relu,relu,linear)\\\hline
\hfil \footnotesize Critic params & \footnotesize 4608 & \footnotesize 4608 & \footnotesize 384 & \footnotesize 650\\\hline
\hfil \footnotesize Total params & \footnotesize 18581 & \footnotesize 18532 & \footnotesize 1636 & \footnotesize 2749\\\hline
\end{tabular}
\end{center}
\caption{Architectures and hyperparameters of the SAC and QSAC configurations.}
\label{table:configurations1}
\end{table*}
\begin{table*}
\begin{center}
\begin{tabular}{|l|c|c|c|c|c|}\hline
\diagbox[width = 2.6 cm, height = 1.2 cm]{\footnotesize \bf Hyperparams}{\footnotesize \bf Model}
&{\begin{minipage}{2.1 cm}\footnotesize \bf SAC\\ 3000 total parameters \end{minipage}}&{\begin{minipage}{2.1 cm}\footnotesize \bf SAC\\ 270000 total parameters \end{minipage}}&{\begin{minipage}{2.1 cm}\footnotesize \bf Full QSAC\\ 2700 total parameters \end{minipage}}\\\hline
\hfil \footnotesize $\gamma$ & \footnotesize 0.99 & \footnotesize 0.99 & \footnotesize 0.99\\\hline
\hfil \footnotesize $\alpha$ & \footnotesize 0.2 & \footnotesize 0.2 & \footnotesize 0.2\\\hline
\hfil \footnotesize Learning rate & \footnotesize 0.0003 & \footnotesize 0.0003 & \footnotesize 0.0003\\\hline
\hfil \footnotesize Memory size & \footnotesize 1000000 & \footnotesize 1000000 & \footnotesize 1000000\\\hline
\hfil \footnotesize Optimizer & \footnotesize Adam & \footnotesize Adam & \footnotesize Adam\\\hline
\hfil \footnotesize Actor neurons & \footnotesize (6,7)(8,(1,1)) & \footnotesize (6,7)(8,(1,1)) & \footnotesize (6,VQA(5 layers),8,1)\\\hline
\hfil \footnotesize Actor act. func & \footnotesize (linear,relu,linear) & \footnotesize (linear,relu,linear) & \footnotesize (linear,relu,linear)\\\hline
\hfil \footnotesize Actor params & \footnotesize 149 & \footnotesize 149 & \footnotesize 112\\\hline
\hfil \footnotesize Critic neurons & \footnotesize (8,22)(22,21)(21,1) & \footnotesize (8,256)(256,256)(256,1) & \footnotesize (8,VQA(20 layers),8,1)\\\hline
\hfil \footnotesize Critic act. func. & \footnotesize (linear,relu,relu,linear) & \footnotesize (linear,relu,relu,linear) & \footnotesize (linear,relu,relu,linear)\\\hline
\hfil \footnotesize Critic params & \footnotesize 719 & \footnotesize 67840 & \footnotesize 650\\\hline
\hfil \footnotesize Total params & \footnotesize 3025 & \footnotesize 271509 & \footnotesize 2712\\\hline
\end{tabular}
\end{center}
\caption{Architectures and hyperparameters of the SAC and Full QSAC configurations.}
\label{table:configurations2}
\end{table*}

%% file: 07_Conclusions.tex
A quantum variational approach to the Soft Actor-Critic algorithm has been studied, looking for and evaluating possible quantum advantages over its classical counterpart when applied the control of a simulated robotic arm. 
Threes types of quantum-classical hybrid network architectures of the Soft Actor-Critic algorithm have been experimented by making use of digital simulation of quantum circuits. Specifically, a SAC with a quantum-classical hybrid Actor network, a SAC with a hybrid Critic networks, and a third algorithm with hybrid Actor and Critic components. 
For all architectures, a comparison with the classical implementation of the  algorithm with comparable number of learnable parameters has been carried out. While no quantum advantage has been found with a hybrid Actor component of the algorithm, a boost in performance has been observed when the SAC algorithm leverages hybrid networks for the Critic components. In particular, the quantum algorithm achieves the same performance as the classical SAC with approximately six times fewer learnable parameters. 
Although the Critic network component seems to be the one primarily affecting performances of the SAC algorithm when applied to robotic arm control, this property gets a novel behavior once quantum circuits are introduced throughout the SAC network components. Indeed, a Full Quantum SAC, with both Actor and Critic components leveraging Variational Quantum Circuits, exhibits a huge quantum advantage with respect to its classical counterpart. Namely, it requires 100 times fewer learnable parameters to reach the same performance as the classical SAC model. The possible reasons of this remarkable reduction in the number of learnable parameters have been argued: the expressiveness of the quantum-classical hybrid Actor component better reproduces the RL agent policy; the Actor, Critic and target Critic components influence each other during the training process, and the better performance of only one of such components may result in a slowdown in the optimization of the learnable parameters.

As a future work, an amplitude encoding of the state action will be tested in order to introduce more non-linearity to  the quantum circuits, possibly avoiding the use of classical dense neural network layers in the Critic and Actor components of the algorithm. Moreover, the evaluation of the quantum algorithm with a physical quantum computer will be carried out. Indeed, while data re-uploading seems to work particularly well for quantum reinforcement learning using simulators, further experiments are required in order to understand if such an approach could be exploited on real quantum devices, dealing with decoherence and noise, and suffering from potential design limitation.